\documentclass{elsart}

\usepackage[latin1]{inputenc}
\usepackage{amsmath}
\usepackage{amsfonts}
\usepackage{amssymb}
\usepackage{graphicx}
\newcommand{\be}{\begin{equation}}
\newcommand{\ee}{\end{equation}}

\begin{document}

\begin{frontmatter}

\title{A logistic map approach to economic cycles \\ I. The best adapted companies }
\author[wroclaw,liege]{J. Mi\'skiewicz},
\ead{jamis@ozi.ar.wroc.pl}
\author[liege]{M. Ausloos}
\ead{marcel.ausloos@ulg.ac.be}

\address[wroclaw]{Department of Physics and Biophysics, University of Agriculture, ul. Norwida 25, 50-375 Wroc\l{}aw, Poland}
\address[liege]{SUPRATECS and GRASP, Institute of Physics, B5, University of Li$\grave e$ge, B-4000 Li$\grave e$ge, Euroland }

\begin{abstract}
A birth-death lattice gas model about the influence of an
environment on the  fitness and concentration evolution of economic entities is
analytically examined. The model can be mapped onto a high order logistic map.
The control parameter is a (scalar) ''business plan''. Conditions  are searched
for growth and decay processes, stable states, upper and lower bounds,
bifurcations, periodic and chaotic solutions. The evolution equation of the
economic population for the best fitted companies indicates ''microscopic
conditions'' for cycling. The evolution of a dynamic exponent is shown as a function of the business plan parameters. 
\end{abstract}

\begin{keyword}
non-equilibrium, birth-death process, external field
\PACS{05.45.-a, 05.45.Ra, 89.65.-s, 89.65.Gh}
\end{keyword}

\end{frontmatter}

\maketitle
\section{Introduction}
Economic cycles (EC) (see $http://cepa.newschool.edu/het/schools/business.htm$)
have been noticed ca. 1930
\cite{cycleKaldor40,cycleKalecki}. They have received
some political economy foundation \cite{cycleLong83,cycleBasu99,4b,8c,cyclehistory,cycleGabisch87}, and are often said to be due to time delayed
competition between ''macroscopic variables'', investment, production, profit, credit conditions,
... basically controlled by the interest rate \cite{cyclehistory,cycleGabisch87,4c}. EC
have not always been well described or defined \cite{cycleKS01,cyclemeasure,4a}.
Cycles are sometimes confused with oscillations, or even noise, in
economic data.  The measurement of the ''period'' of a cycle is even far from
what would be expected in standard physical measurements.

Economists admit that EC periods are ill defined \cite{cycleKS01,cyclemeasure,4a} and do depend on what goods, assets, ... are considered. Following averaging procedures one can consider some constancy (3-4 years) in some EC like from the 1994 NBER table. Sometimes not constant periods are said to be the rule, and their values checked for according to various data analysis techniques \cite{8a,8b}.

EC have recently received some renewed attention in the econophysics community
\cite{cycleAoki01,gabaix,weron,8e}. One approach that can be thought of is through 
the stochastic resonance phenomenology in which
noise and (or) shocks might lead and/or induced a stability state, or some
dissipative structure \cite{Glansdorf}, smoothly evolving  \cite{Nicolis
Nicolis}, - an oscillating trend. A modern econophysics approach would  also bear
upon self-organization \cite{Bak,Sornette}, a notion already thought of by
economists \cite{SocecoFoster}.

Many analytical and simulation approaches are based on the Kaldor -- Kalecki model \cite{cycleKaldor40,cycleKalecki} or generalizations like in \cite{4b,16a}. With as many parameters one can write coupled Langevin equations for fluxes or rates \cite{16b}. The main difficulties rests in measuring the coupling parameters.

A simulation with cellular automata or neural networks might lead to a 
description of  EC, but there are many caveat in these approaches since they also
contain  parameters or black boxes which are either ad hoc ingredients or  simply
far from so called ''first principles''.

Another modern approach implies a microscopic-like description for such a macroeconomic problem. This can be at departure from usual continuous time evolution equations but the less so if some algorithm is analytically rewritten as here below.  

Our analysis stems from  a numerically analyzed model  \cite{ek:aus-pek}
considering the evolution of a set of entities (agents or companies) under
changing economic, or more generally environmental, conditions. A question raised
was in fact whether the system can have bifurcation points. The model 
is a lattice gas in which {\it mobile} entities are
described by a scalar variable degree of freedom, itself compared to what is
called a field \cite{ek:aus-pek}, - controlling the most probable state of the
system. The degree of
freedom evolution depends also on the environment of the entity. The economic
evolution of the entities is imposed to have a Lamarckian feature  as often
admitted in economy circles \cite{Lamarckeco}.  A Verhulst attrition-like term is
inserted through the company business plan which serves for outlining the
birth-death process. A statistical physics approach based on the logistic map seems thus suitable for describing so called economic-like cycles. {\it A posteriori} one might wonder why it  has not been used, since the
logistic map ingredients, in the original Verhulst work, stems indeed  from
socioeconomic considerations \cite{Schuster}. Recall that a pedagogical example
of a logistic map  application is the interest rate effect on savings account
\cite{Schuster}. It will be found that such ''microscopic'' ingredients 
lead to stable states, bifurcations, periodic and
chaotic (turbulent-like) regimes. Therefore it is pointed out that so called EC
belong to a class of self-organized systems and  may have simply controllable
features. Quite importantly it seems that the macroscopic ingredients of usual EC
theories can be replaced by more microscopic inputs. We are of course aware
of the present simplicity of the model. EC are not due to a birth and death process only.

\section{ACP model}
Let us precise the basic ingredients of the model  \cite{ek:aus-pek}, as used
here. A set of entities (called ''companies'') is originally placed at random on
a (square symmetry) lattice. Each company $ i  $ is characterized by a real,
scalar degree of freedom  $f_i$ $\in [0,1] $. This parameter denotes how well a
company is adapted to the economic environment, itself characterized by a (real)
field  $ F \in [0,1] $. The system is e.g. described by the concentration of
companies, $ c_t \in [0,1] $. The concentration is defined in a usual way $ c_t=
\frac{N_{tot}(t)}{N_{max}} $, where $ N_{tot}(t) $ is the number of companies on
the lattice at time $ t $ and $ N_{max} $ is the volume (the number of sites on
the lattice) of the system. The time $ t $ is discrete and counted in Monte Carlo
steps (MCS). The companies may diffuse one lattice spacing at a time. Although in the
original model \cite{ek:aus-pek} the field was space and time
dependent, it will be taken constant in space and time here. Unlike in
\cite{ek:aus-pek}, where the system was described by the position and state of
each individual company, within this paper, the system is described through a
{\it distribution function} $ N(t,f) $ -- the number of companies characterized
by $ f  $ at time  $ t $. In this case the total number of firms existing in the
environment at time $ t $ is given by: 

\begin{equation} 
\label{total} 
N_{tot} (t) = \int_{0}^{1} N (t,f) df 
\end{equation}

In one MCS a number of companies  equal to that at the
beginning of the step is picked at random and possibly displaced by one lattice
site. After each move the following changes are
allowed, if possible: 
\begin{enumerate}
\item The company may disappear due to
"difficult economic conditions" as measured by $ F $; the probability of
surviving is 
\begin{equation} 
\label{prob} 
p_i = \exp(-sel |f_i -F|) ,
\end{equation} 
where $ sel $ is a positive parameter describing the "selection
pressure". This means that the largest survival probability is when the
enterprise $ i $ satisfies the condition: $ f_i = F $; it is a  {\it best adapted
company};
\item companies may merge. If company $ i $ has a nearest
neighbor $j$, either with a probability $ b $ both form a new company $k$  on 
site $i$ while $ j $ disappears with \\
\begin{equation} 
\label{merg} 
f_k = \frac{f_i +f_j}{2} +  sign[0.5 -r]
\frac{|f_i- f_j|}{2}, 
\end{equation} 
where $ r $ is a random number, $ r \in
[0,1] $; 
\item or with a probability $ 1-b $, companies
{\it survive and} create one (or two) new firm(s), depending on the available space
in the Moore neighborhood of the $ i $--company, each with a specific random $f$.
\end{enumerate} 

\section{Mean field approximation}
Averaging over a MCS such that 
$$
<\frac{f_i +f_j}{2} +  sign[0.5 -r] \frac{|f_i - f_j|}{2}> 
=   <\frac{f_i +f_j}{2}>   \rightarrow   f_k (t+1)
$$
taking into account the modification in the state distribution caused by (ii), i.e.
 $ N (t+1,f) \leftarrow N (t,(f+F)/2)$ and
considering all possible events in a mean field sense the evolution 
equation of the system can  be
written as the sum of two terms, 
\begin{equation} 
\label{evol1} N (t+1,f) =
H_1(c_t) p(f) N (t,f) + H_2(c_t) N (t,g(f)) , 
\end{equation} 
where (dropping the
$t$ index in $c_t$) : 

\begin{eqnarray}
H_1 (c) & = & 1-b (1-c)[4c^3 + 6c^2 (1-c) + 4c(1-c)^2 ] \nonumber \\ 
& & + 4(1-b)(1-c)c^7 \\
& & + 2(1-b)(1-c) \{ 1-c^7 - 4(1-c)^6c \nonumber \\ 
& & \;\;\; - 6(1-c)^5c^2  - 4(1-c)^4 c^3
-(1-c)^3c^4] \} , \nonumber \\
H_2 (c)  & = & b(1-c)[4c^3 + 6c^2 (1-c) +4c(1-c)^2 ] 
\end{eqnarray} 
 and 
\be
g(f) =\frac{f+F}{2}. 
\ee

Eq.(\ref{evol1}) can be iterated back until the initial point
\begin{equation} 
\label{evol2} 
N (t,f) = \sum_{i=0}^t a_{i,t-i} p^{t-i} (f) N
(0,g^i (f)),   
\end{equation} 
where each coefficient $ a_{i,t-i} $ is the sum of time
oriented products of $ H_1(c_t) $ and $ H_2(c_t) $, i.e.  
\be
 a_{i,t-i}  =
\sum_\sigma H_{\sigma (t) } (c_{t_0}) H_{\sigma (t-1) } (c_{t_1})   \ldots
H_{\sigma (t-i) } (c_{t_{i}}). 
\ee
 where $ t_0> t_1> t_2> \ldots > t_{i} $ and $
\sigma  $ denotes all possible permutations of the set $ \{
\underbrace{1,1,\ldots,1}_{i \mbox{times}},\underbrace{2,2,\ldots,2}_{t-i \mbox{
times}} \} $.

\section{Logistic map}
\begin{figure} 
\begin{center}
\includegraphics[scale=0.35,angle=-90]{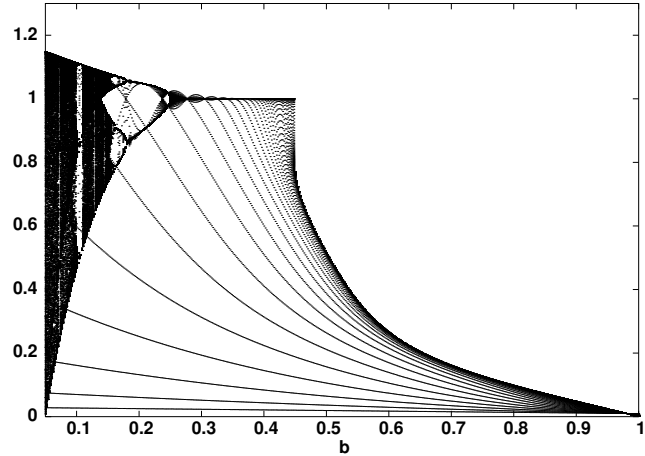}
\end{center}
\caption{ \label{fig_log} Iterations of the concentration as a function of
the parameter $b$ starting from a low initial concentration}
\end{figure}

\begin{figure} 
\begin{center}
\includegraphics[scale=0.35,angle=-90]{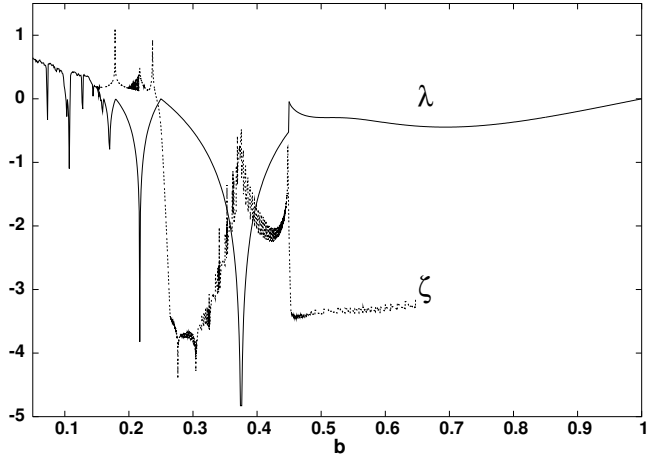} 
\end{center}
\caption{\label{fig_lap} Lyapunov ($\lambda$) (solid line) and the
dynamic exponent $ \zeta $ (dashed line) as a function of $b$}
\end{figure}

The evolution equation for the total number of the best adapted companies has the form of a
logistic map \cite{Schuster,ek:map}, with a control parameter $ b $.
Therefore the model may reveal some
periodic or chaotic behavior. The best adapted
company concentration evolution, starting from a small initial
value, is shown at fixed $b$'s in Fig. \ref{fig_log} for different $ b $ values.
The corresponding Lyapunov exponent $\lambda$ is shown in (Fig. \ref{fig_lap}).
Five main regions can be distinguished in Fig. \ref{fig_lap}. Only one stable
solution (with $ c_t > 1 $) exists for $ b>0.45 $, i.e. due to the frequent merging process, there
is no overpopulation of the system. The stable state is achieved as a dynamic 
equilibrium between
merging and spin-off creation process. For other $b$ values some "convergence" is
found toward one or several states characterized by bifurcations and limit cycles
 with various periodicity down to $b_c \simeq 0.1556$, where a chaotic
region appears containing  narrow stability windows.

\begin{figure} 
\begin{center}
\includegraphics[scale=0.35,angle=-90]{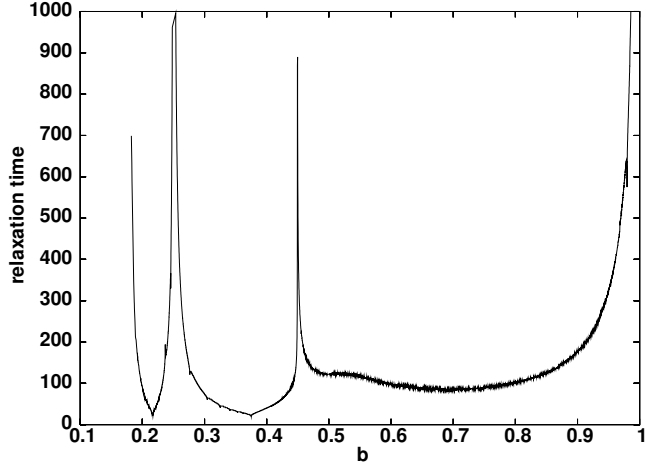}
\end{center}
\caption{\label{fig_spek} Time for achieving a stable state as a function of b } 
\end{figure}

\begin{figure} 
\begin{center}
\includegraphics[scale=0.35,angle=-90]{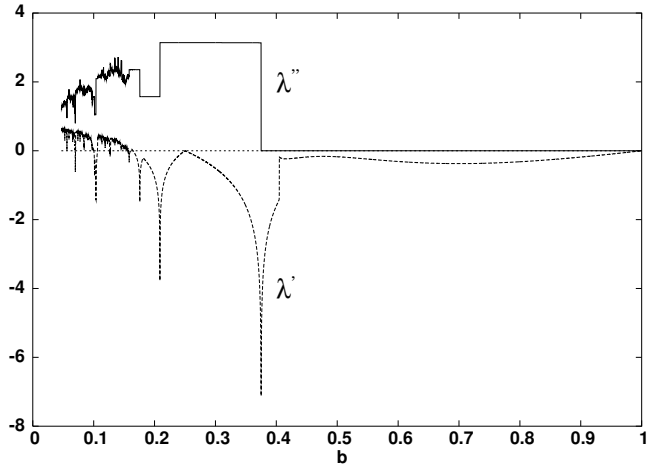}
\end{center}
\caption{\label{fig_img} Real and imaginary part of a Lyapunov-like exponent }
\end{figure}

In order to better understand the dynamics we calculated 
\begin{enumerate}
\item the time needed
before a stable orbit (state) is reached, Fig. \ref{fig_spek}, and
\item a variant
of the Lyapunov exponent in order to emphasize both the convergence (or
divergence) of the trajectory, $and$ a possible oscillatory behavior, through the
so called real ($\lambda'$) and imaginary ($\lambda''$) part of a ''generalized''
$\lambda = ln [(d/dx) f(x)]$ in conventional notations, but without absolute
values  before taking the logarithm, Fig. \ref{fig_img}.
\end{enumerate} 

For $ b>0.45 $ the time required for the system to attain the stable state
remains (approximately) on the same level until $ b<0.65 $. However for $
b\approx 1 $ the time required for the system to reach the stable state increases
rapidly (Fig. \ref{fig_spek}). This corresponds to the behavior of the Lyapunov exponent
(Fig. \ref{fig_lap}), which is $ \lambda = 0 $ for $ b=1 $. Especially interesting
is the range $ b \in (0.38;0,45) $: damping properties are superposed to an
oscillating behavior. Oscillations are also visible in Fig. \ref{fig_img}, where
the finite imaginary part of  $\lambda$ shows that the system may cycle.   Notice
the value of $\lambda ''$ equal to $\pi$, $\pi/2$, $3\pi/4$,... in various
stability regions, but taking bizarre values in chaotic regions. This is another
way to find characteristics of stability windows in the "turbulent regimes".

The process of achieving a stable state has been also
studied by assuming a power law decay for the $c$-distance evolution in
the form $ |\dot{c}(t) | \sim t^\zeta $. The plot of the $ \zeta $ exponent as a function of
$ b $ is shown in Fig. \ref{fig_lap}. 

\section{Conclusions}
The above results, for a simple model, lead toward interesting conclusions
concerning stable or not economic states. 
We do not - of course claim that EC are due to a merging -- spin off microscopic process only. What is intended is to search whether a discrete microscopic model has some interesting ingredients for some macroeconomic evolution.
The evolution of the model
depends on the three microscopic-like parameters the environmental condition $ F
$, the selection pressure $ sel $ and the merging probability $ b $ having a physical sense in contrast
to macroscopic (economic) ones . The
influence of the environment $ F $ and the selection
pressure control the time required for the system to achieve a
stable state. Moreover the environmental condition decides also on the type of
companies surviving in the system. The first parameter is essentially of
political and/or global origin. The second can be taken as the inverse of
temperature in thermodynamics, and leads as in other non equilibrium systems to
consider an intrinsic noise controlling the existence of dissipative structures
\cite{Glansdorf,Schuster}. If the selection pressure is ''strong'' then the
"weak" companies are removed from the system and the distribution function $
N(t,f) $ converges to the "best adapted companies" case $ N(t,F) $. This is also similar to
results found in Darwinian-like economic evolutions \cite{KYamasaki,Darwineco}.
The last parameter $b$ is more subtle, and concerns business plans: it controls
the asymptotic concentration value  of companies and the cyclic or chaotic
behavior of the system.
We have constrained $ b $ to be a constant though a dynamic economic policy usually requests a time dependent $ b $.

Concentrating on the best adapted companies for small $ b $ the system reveals
chaotic properties, i.e. $\lambda > 0 $. It means that at
low merging the system becomes unstable. This is an apparently new interesting
feature which seems reasonable but nevertheless requests economic studies and
considerations. The system is especially stable at $ b\approx 0.38 $ where the
damping parameter attains its largest value. In such a case, the system has only
one stable solution, which is achieved in the shortest possible time. Therefore a
merging philosophy,  or politics, is a stabilization factor for economic systems.
$In$ $fine$, it should leads to a monopolistic situation which is stable and
''cycling'' as long as the goods of such a company are needed, - if there is no
spin-off process.

Finally, the above numbers and findings should truly be taken with caution since they pertain to a model studied on a 2D lattice
with a given symmetry. The period of cycles depends on the economic plan 
parameter $b$, but also on the Moore
neighborhood hereby chosen for the evolution and the information flow between
companies. {\it A posteriori } it is  understood why ''EC periods'', either
global or for specific activity fields, are poorly defined \cite{cycleLong83,cycleBasu99,4a}. Interestingly it
appears again that the connectivity of the ''lattice'' on which an economic
process takes place is markedly relevant for defining economic oscillations. This
has already been noticed in the case of financial crashes \cite{Sornette,Tokyo2}.
Whence some  EC appear to be self-organized systems and have simply controlled (and
controllable) features, similar to turbulent processes.

\section{Acknowledgments} This work is partially financially supported by FNRS
convention FRFC 2.4590.01. J. M. would like also to thank GRASP
for the welcome and hospitality.

\end{document}